\documentclass[12pt,a4paper]{article}

\usepackage{graphics}
\usepackage{latexsym}
\usepackage{amsmath}
\usepackage{amssymb}

\title{Trying to understand mass}
\author{B. Hoeneisen}
\date{\small{
1 September 2006} }

\begin{document}
\maketitle

\begin{abstract}
\noindent
We try to understand how particles acquire
mass in general, and in particular, how they acquire mass
in the standard model and beyond. 
\end{abstract}


\section{Introduction}
At the end of my talk at the Galapagos
World Summit on Physics Beyond the Standard 
Model\footnote{San Cristobal Island, 22 -- 25 June 2006.}
I mentioned, in passing, three crazy ideas.\footnote{Ideas 
developed incompletely, so that their usefullness
(or correctness) has not been established.}
In this talk I elaborate on one of them.

The origin of mass is a mystery.
Particles may acquire mass, or change their mass, when they
interact with a medium. Examples are photons interacting
with free electrons in a plasma or in a metal,
photons interacting with Cooper pairs in a superconductor, 
electrons propagating in the periodic potential of
a crystal, and photons propagating in a waveguide.

The symmetries of the standard model Lagrangian prevent adding
\textquotedblleft{by hand}" mass terms for the fermions or bosons. 
These particles acquire mass due to their interaction
with the Higgs field.

The limit $m \rightarrow 0$ is singular
in the sense that phenomena present with $m \ne 0$ is absent
with $m = 0$. Examples are the longitudinal polarization
of $W^+$, $W^-$ and $Z$, the coupling between the left and right
Weyl components of the Dirac field, and the 
Cabibbo-Kobayashi-Maskawa matrix.

The origin of mass is the next frontier in high energy physics.
In this talk I try to understand how particles acquire
mass in general, and in particular, how they acquire mass
in the standard model and beyond. 

\section{Dispersion relation}
Let us consider the wave
\begin{equation}
\propto e^{ i (\vec{k} \cdot \vec{r} - \omega t) }
\label{wave}
\end{equation}
with the dispersion relation
\begin{equation}
\omega_0^2 = \omega^2 - k^2 c^2.
\label{omega_k}
\end{equation}
$\omega_0$ and $c$ are constants.
Let us give some examples. 
For an electromagnetic
wave in vacuum, $\omega_0 = 0$ and $c$ is the
velocity of light. 
For a sound wave, $\omega_0 = 0$ 
and $c$ is a velocity of sound. 
For an electromagnetic 
wave in a plasma or in a metal, 
$\omega_0 = \left[ e^2 n /(\epsilon_0 m) \right]^\frac{1}{2}$ is the
plasma frequency and $c$ is the
velocity of light. 
For an electromagnetic wave in
a waveguide, $\omega_0$ is the cut-off frequency of
the mode of propagation and $c$ is the
velocity of light.
For an electromagnetic wave in a dielectric,
$\omega_0 = 0$ and $c$ is the
velocity of light in the dielectric.
For a Dirac (spin-$\frac{1}{2}$) particle,
$\omega_0 = m c^2 / \hbar$ and $c$ is the
velocity of light.

To every wave (\ref{wave}) there is associated a
particle of energy 
\begin{equation}
E = \hbar \omega
\label{E_omega}
\end{equation}
(Planck relation), momentum
\begin{equation}
\vec{p} = \hbar \vec{k}
\label{p_k}
\end{equation}
(De Broglie relation), and velocity equal
to the group velocity of the wave:
\begin{equation}
\vec{v} = \nabla_{\vec{k}}(\omega)
  = \frac{\vec{k} c^2}{\omega}
  = \frac{\vec{p} c^2}{E}.
\label{v}
\end{equation}
From the preceding equations we obtain the dispersion
relation of the particle,
\begin{equation}
m^2 c^4 = E^2 - p^2 c^2,
\label{E_p}
\end{equation}
where the \textquotedblleft{cut-off mass}"
is given by
\begin{equation}
m c^2 = \hbar \omega_0.
\label{omega0_m}
\end{equation}
If the particle has zero mass,
\begin{equation}
E = pc, \qquad \omega = kc.
\label{k_omega}
\end{equation}
If the particle is massive, we obtain
\begin{equation}
E = \frac{m c^2}{\sqrt{1 - (v/c)^2}},
\label{E_v}
\end{equation}
\begin{equation}
\vec{p} = \frac{m \vec{v}}{\sqrt{1 - (v/c)^2}}.
\label{p_v}
\end{equation}

For a particle in vacuum, $c$ is the velocity of
light, and the energy-momentum $(E, \vec{p})$,
given by (\ref{E_v}) and (\ref{p_v}), transforms
the same as the space-time coordinates $(t, \vec{r})$.
Therefore, the energy-momentum of a particle in vacuum
is a 4-vector with respect to the Lorentz group.

\section{Inertia}
Is the \textquotedblleft{cut-off mass}" related to
the \textquotedblleft{inertial mass}" in Newton's
equation? Yes! Let us give an example. Consider a 
horn, i.e. a waveguide with transverse dimensions
$(x, y)$ increasing with $z$. 
The group velocity $v_z$ increases with $z$,
the corresponding particle is accelerated, and the
particle momentum increases as
\begin{equation}
F \equiv \frac{dp_z}{dt} = \frac{d}{dt} \left( \frac{Ev_z}{c^2} \right) 
= \frac{E}{c^2} \frac{dv_z}{dt}
\approx m \frac{dv_z}{dt}
\label{Newton}
\end{equation}
if $v \ll c$. Therefore, the \textquotedblleft{inertial mass}" is equal to
the \textquotedblleft{cut-off mass}" $m$.
$F$ is the force.
The horn momentum increases by the same amount per unit
time in the opposite direction. $E = \hbar \omega$ is constant 
because the horn does not change with time.
$p_z$ is not constant because the horn varies with $z$.

In general, consider any system with energy $E_0$ in its
rest frame, \textit{i.e.} the frame with $\vec{p} = 0$.
After a Lorentz transformation,
the system has velocity $\vec{V}$ and
momentum $\vec{p} = E_0 \vec{V}/c^2$ if $V \ll c$. 
So the system has
inertia with mass $m = E_0 / c^2$. Inertia, and 
the conservation of energy-momentum, are two aspects of
the same phenomenon.

\section{Conservation of energy-momentum}
Let us try to understand the conservation of 
energy-momentum. As an example,
we consider this experiment: 
we illuminate an atom with two beams
of light, and observe the light scattered
in the direction $\vec{k}_3$ with a
receiver far away.
The atom is at $\vec{r}_j$, and the 
receiver is at $\vec{r}$.
The electric field of the incident 
beams at the atom is
\begin{equation}
  A_1 e^{i ( \vec{k_1} \cdot \vec{r_j} - \omega_1 t )} 
        + A_2 e^{i ( \vec{k_2} \cdot \vec{r_j} - \omega_2 t )}.
\label{sum}
\end{equation}
This field induces dipole, quadrupole, etc, moments
in the atom, so the atom emits an electromagnetic
wave. The amplitude of this scattered wave
in the direction $\vec{k}_3$ is some
non-linear function of (\ref{sum}).
Expanding this non-linear function in a Taylor
series, we find that the amplitude of the
scattered wave has the form
\begin{equation}
\sum_{n, m} { a_{nm} e^{i n (\vec{k_1} \cdot \vec{r_j} - \omega_1 t )} 
e^{i m ( \vec{k_2} \cdot \vec{r_j} - \omega_2 t )}},
\label{scattered}
\end{equation}
with $n, m = 0, \pm 1, \pm 2 \cdots$.
The electric field at the receiver is proportional to
\begin{equation}
\propto \sum_{n, m}{ a_{nm} e^{i n (\vec{k}_1 \cdot \vec{r}_j - \omega_1 t )}
e^{i m ( \vec{k}_2 \cdot \vec{r}_j - \omega_2 t )}
e^{i \vec{k}_3 \cdot ( \vec{r} - \vec{r}_j )}}.
\label{receiver}
\end{equation}
Note that the amplitude at the receiver has
components of frequency
\begin{equation}
\omega_3 = n \omega_1 + m \omega_2.
\label{omega}
\end{equation}
Multiplying by $\hbar$ (to get
the conventional unit for energy)
we obtain the equation of conservation
of energy:
\begin{equation}
\hbar \omega_3 = n \hbar \omega_1 + m \hbar \omega_2.
\label{energy}
\end{equation}
In our example with positive $n$ and $m$, $n$ 
incoming photons of energy
$E_1 = \hbar \omega_1$ scatter on $m$ incoming photons of 
energy $E_2 = \hbar \omega_2$, producing an outgoing photon of
energy $E_3 = \hbar \omega_3$. Note that
\textit{classical} waves interact in packets of
frequency or energy.

It is convenient to include a factor 
$e^{i ( \vec{k}_3 \cdot \vec{r} - \omega_3 t)}$
in the proportionality constant of (\ref{receiver}).
Then equation (\ref{receiver}) becomes
\begin{equation}
\propto \sum_{n, m} { a_{nm} e^{i n (\vec{k}_1 \cdot \vec{r}_j - \omega_1 t )}
e^{i m ( \vec{k}_2 \cdot \vec{r}_j - \omega_2 t )}
e^{-i ( \vec{k}_3 \cdot \vec{r}_j - \omega_3 t)}}.
\label{receiver2}
\end{equation}
The part exhibited explicitly in (\ref{receiver2})
is independent of $(t, \vec{r})$.

Let us now consider a crystal. The amplitude at the
receiver is a sum of terms (\ref{receiver2})
over all atoms $j$ in the crystal:
\begin{equation}
\propto \sum_{j, n, m} { a_{nm} e^{i n (\vec{k}_1 \cdot \vec{r}_j - \omega_1 t )}
e^{i m ( \vec{k}_2 \cdot \vec{r}_j - \omega_2 t )}
e^{-i ( \vec{k}_3 \cdot \vec{r}_j - \omega_3 t)}}.
\label{wave_out}
\end{equation}
Since, for a 3-dimensional crystal lattice,
$\vec{r_j} = \vec{r_0} + N \vec{a} + M \vec{b} + L \vec{c}$
with $N$, $M$ and $L$ integers,
it follows that the condition that the waves 
scattered at each atom $j$ add up in phase at
the receiver is
\begin{equation}
n \vec{k_1} + m \vec{k_2} = \vec{k_3} + \vec{G},
\label{Bragg}
\end{equation}
where $\vec{G} \cdot \vec{a}$, $\vec{G} \cdot \vec{b}$
and $\vec{G} \cdot \vec{c}$ are multiples of $2 \pi$.
Equation (\ref{Bragg}) is the Bragg law,
or, multiplying by $\hbar$, the law of conservation of 
\textquotedblleft{crystal momentum}". $\vec{G}$
is a vector of the \textquotedblleft{reciprocal lattice}". 
$\hbar G$ is the momentum acquired by the crystal as a whole.
In the limit
of a continuous medium, i.e. in the limit 
$\vec{a}, \vec{b}, \vec{c} \rightarrow 0$, the only 
finite $\vec{G}$ is $\vec{G} = 0$, and we
obtain the law of conservation of momentum.

In conclusion, the conservation
of frequency-wave vector (or equivalently, 
energy-momentum) is the condition for the
waves scattered at different space-time points
to add up in phase at the receiver, and is valid
if space-time is homogeneous, i.e. if the vertex
factor in the Lagrangian does not depend 
explicitly on space and time. 
Note that, for scattering to take place, the
Lagrangian must be non-linear, i.e., the vertex
terms must contain 3 or more fields.
Since the powers in the Taylor expansion,
$n, m, \cdots$, are integers, the
interaction occurs in packets 
of energy-momentum 
$(E_i, \vec{p}_i) = (\hbar \omega_i, \hbar \vec{k}_i)$
we call \textquotedblleft{particles}". Note
that the concept of \textit{particle}, 
and the Planck and De Broglie
relations, have emerged from the interaction 
of \textit{classical} waves!
Quantum mechanics tells us that 
nature \textit{chooses} among the various scatterings
$n, m$ with probabilities proportional
to $|a_{nm}|^2$.

Note that the Feynman rules to obtain
the probability amplitude for a particular scattering is
obtained by multiplying a factor 
$\exp { [ i ( \vec{k_i} \cdot \vec{r_j} - \omega_i t) ] }$
for each incoming particle $i$, a factor
$\exp { [ -i ( \vec{k_o} \cdot \vec{r_j} - \omega_o t) ] }$
for each outgoing particle $o$,
and a vertex factor proportional
to $a_{nm}$.

The example can be generalized to an arbitrary number
of sources and receivers (or, equivalently, an arbitrary
number of incoming and outgoing particles). If we
consider re-scatterings, we arrive at 
Feynman's sum over paths.

Let us mention two interesting points:
\begin{itemize}
\item
The accuracy with which the conservation of
energy-momentum is valid depends on the size of the 
interaction region and the time it takes for
the interference to build up. The result of a
\textquotedblleft{back-of-the envelope}" calculation
is the \textquotedblleft{uncertainty principle}".
\item
Energy-momentum is conserved in the 
reference frame in which the
\textquotedblleft{crystal}" is at rest.
Since, in vacuum, energy-momentum transforms as a
4-vector, energy-momentum is also conserved in a
reference frame with constant velocity with respect
to the \textquotedblleft{crystal}".
\end{itemize}

In summary, we have gained some insight into
why energy-momentum is conserved,
and why interactions occur in \textit{packets}
satisfying the Planck and De Broglie relations.

\section{Particle in a box}
Consider a
particle that collides elastically with the sides of 
a box.\footnote{This model was inspired by \cite{Wilczek}.}
The box is accelerated in the $z$-direction by some
external means. At time $t=0$ the box is at rest (with respect
to the inertial reference frame), and the particle collides
elastically with one wall of the box. At time $t = \Delta t$
the particle collides with the opposite wall of the box, which
is now moving in the $z$-direction with velocity 
$V = a \cdot \Delta t$.
The change of $p_z$ of the particle is less in
the second collision than in the first collision due to the
velocity $V$. The difference, calculated doing two 
Lorentz transformations, is
\begin{equation}
\Delta p = \frac{2V}{c^2} E
\label{Dp}
\end{equation}
if $V/c \ll 1$. The change of momentum per unit time is
called \textquotedblleft{force}". The time to go back and
forth across the box is $2 \Delta t$ if $V/c \ll 1$.
Therefore the force is given by Newton's equation
\begin{equation}
F = M \cdot a
\label{F}
\end{equation}
with Einstein's famous inertial mass
\begin{equation}
M = \frac{E}{c^2}.
\label{M}
\end{equation}
This relation is valid for particles in the box
of arbitrary velocity:
the energy per particle is 
(\ref{k_omega}) if $m = 0$, or
(\ref{E_v}) if $m > 0$.

It is instructive to repeat this problem with
waves, applying appropriate boundary conditions
at the moving wall. The same result is obtained.

\section{How does the proton acquire mass?}
The proton is composed of two $u$
quarks and one $d$ quark tied
together by gluons.
\footnote{This crude toy model neglects quarks in the sea,
and the energy of gluons.}
The mass of the proton is much larger than the mass
of the $u$ or $d$ quarks, so we neglect the quark masses.
The gluon force becomes weak when the quarks are close together,
and becomes strong when the quarks are separated by more than
$\approx 2$ fm. This \textquotedblleft{running coupling"
of the gluons determines the radius of the proton 
(which is measured, by photon scattering, to be $1.2$ fm).

We try the following crude model: the
proton is composed of three massless quarks in a box.
\footnote{Equivalently, the proton
may be viewed as a bubble in superconducting vacuum at pressure
$P$ filled with an ultra relativistic degenerate Fermi gas with 3
quarks. (?)}
For simplicity we take the box to be a cube of sides $d$ with the same volume
as a sphere of radius $1.2$ fm. We place each of the three quarks
in the lowest energy state with $\lambda / 2 = d$. 
Then we obtain an inertial mass of
the proton of order $\approx 0.96$ GeV$/c^2$.

Excited states are predicted and observed.
\footnote{It is best to use spherical coordinates 
to classify the excited states by their quantum numbers.}

\section{Exploring the limit $m \rightarrow 0$.}
From now on we generally set $c = \hbar =1$.
A massive free field $\phi$ satisfies the wave equation
\begin{equation}
\partial_\mu \partial^\mu \phi + m^2 \phi = 0,
\label{wave_equation}
\end{equation}
which is equivalent to the dispersion relation (\ref{omega_k}).
This differential equation can be written in integral form as
\begin{equation}
\phi (x) = \phi_0 (x) - \int S_0 (x - y) m^2 \phi (y) d^4 y,
\label{integral_equation}
\end{equation}
where $\phi_0 (x)$ is the general solution of
\begin{equation}
\partial_\mu \partial^\mu \phi_0 = 0,
\label{phi_0}
\end{equation}
and $S_0(x)$ is a particular solution of 
\begin{equation}
\partial_\mu \partial^\mu S_0 (x) = \delta^4 (x).
\label{S_0}
\end{equation}
A solution of (\ref{S_0}) is
\begin{equation}
S_0(x) = -\frac{1}{(2 \pi)^4} \int { \frac{e^{-ikx}}{k^2} d^4k }.
\label{solution}
\end{equation}
The $dk^0$ integral can be done (using the Cauchy theorem
with the Feynman handling of the poles) and we obtain
\begin{eqnarray}
S_0(x) &=& \frac{i}{16 \pi^3} \int {\frac
{\exp [ {i(\vec{k} \cdot \vec{r} - \left| \vec{k} \right| t)} ] }
{\left| \vec{k} \right|} d^3 \vec{k}} \qquad
\textrm{for } t > 0\textrm{, and} 
\nonumber \\
S_0(x) &=& \frac{i}{16 \pi^3} \int {\frac
{\exp [ {i(\vec{k} \cdot \vec{r} + \left| \vec{k} \right| t)} ] }
{\left| \vec{k} \right|} d^3 \vec{k}} \qquad
\textrm{for } t < 0.
\label{Solution}
\end{eqnarray}
The subscript $0$ on $S_0(x)$ reminds us that this is the propagator
of a zero mass field.
Now we can treat $m^2$ as a perturbation, and solve 
(\ref{integral_equation}) iteratively:
\begin{eqnarray}
\phi (x) &=& \phi_0 (x) + \int S_0 (x - y) (-m^2) \phi_0 (y) d^4 y \nonumber \\
& &  + \int S_0 (x - y) (-m^2) S_0 (y - z) (-m^2) \phi_0 (z) d^4 y d^4 z
  + \cdots
\label{series}
\end{eqnarray}
This series can be represented with Feynman diagrams as in Figure
\ref{Feynman}. 
Calculating the terms in the series we obtain
\begin{equation}
\exp [ i ( \vec{k} \cdot \vec{r} - \sqrt{m^2 + |\vec{k}|^2} t ) ]
= \exp [ i (\vec{k} \cdot \vec{r} - |\vec{k}| t ) ]
\left( 1 - \frac{i}{2} \frac{m^2}{|\vec{k}|} t - 
\frac{1}{8} \frac{m^4}{|\vec{k}|^2} t^2 + \cdots \right),
\label{series2}
\end{equation}
which is an expansion of the solution of (\ref{wave_equation}) around the
solution of (\ref{phi_0}) valid for $m^2/|\vec{k}|^2 \ll 1$.

The scatterings illustrated in Figure \ref{Feynman} conserve energy-momentum
because the scattering amplitude $-m^2$ does not depend on space-time.
Note that the propagation of a massive particle can be viewed
as the propagation of a massless particle that forward scatters
$0, 1, 2 \cdots$ times on the vacuum with amplitude $-m^2$.
We will call this the \textquotedblleft{stepping stone}" model
of mass.

\begin{figure}
\begin{center}
\scalebox{0.8}
{\includegraphics{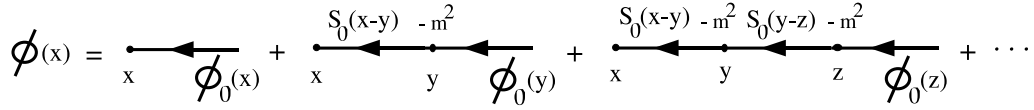}}
\caption{Feynman diagrams of a massless scalar particle
acquiring mass by forward scattering on the vacuum.}
\label{Feynman}
\end{center}
\end{figure}

\section{How does the electron acquire mass?}
The Dirac field $\psi$ has 4 components and carries a reducible 
representation of the proper Lorentz group. The
irreducible components are Weyl-L and Weyl-R 
\footnote{In this article we will call
\textquotedblleft{Weyl-L}" (\textquotedblleft{Weyl-R}")
the representation $\left( \frac{1}{2}, \frac{3}{2} \right)$
($\left( \frac{1}{2}, -\frac{3}{2} \right)$) 
of the proper Lorentz group in the notation of \cite{GMS}.
In this notation, the scalar representation is
$\left( 0, 1 \right)$, and
the vector representation is
$\left( 0, 2 \right)$.}
of dimension 2:
\begin{equation}
\psi = \psi_L \oplus \psi_R, \qquad
\psi_L = \frac{1}{2} ( 1 - \gamma^5 ) \psi, \qquad
\psi_R = \frac{1}{2} ( 1 + \gamma^5 ) \psi.
\label{Weyl}
\end{equation}
The Dirac field carries irreducible representations of
the complete (including space reflection) and general
(including space and time reflections) Lorentz groups.
In other words, proper Lorentz transformations do not mix
$\psi_L$ with $\psi_R$, but space or time reflections do.
The Dirac equation for a free electron can be written as
\begin{eqnarray}
-i \gamma^\mu \partial_\mu \psi_L & = & -m \psi_R, \nonumber \\
-i \gamma^\mu \partial_\mu \psi_R & = &-m \psi_L,
\label{Dirac}
\end{eqnarray}
where $\gamma^\mu$ are the Dirac matrices.

To try to understand how the particle acquires mass we proceed
as follows.
The differential 
equations (\ref{Dirac}) can be written in integral form:
\begin{eqnarray}
\psi_L(x) = \psi_{L0}(x) +
  \int{ [ -i S'_0 (x - y) ] ( -im ) \psi_R(y) d^4y }, \nonumber \\
\psi_R(x) = \psi_{R0}(x) +
  \int{ [ -i S'_0 (x - y) ] ( -im ) \psi_L(y) d^4y },
\label{int_Dirac}
\end{eqnarray}
where
\begin{equation}
-i \gamma^\mu \partial_\mu \psi_{L0} = 0, \qquad
-i \gamma^\mu \partial_\mu \psi_{R0} = 0,
\label{free}
\end{equation}
and
\begin{equation}
-i \gamma^\mu \partial_\mu S'_0(x) = \delta^4 (x).
\label{non_homo}
\end{equation}
Equations (\ref{int_Dirac}) can be iterated and we obtain
the series
\begin{eqnarray}
\lefteqn{ \psi_L(x) = \psi_{L0}(x) +
  \int { [ -i S'_0(x-y) ] (-im) \psi_{R0}(y) d^4y } } \nonumber \\ & &
  + \int { [ -i S'_0(x-y) ] (-im) [ -i S'_0(y-z) ] (-im) \psi_{L0}(z) d^4y d^4z} 
  + \cdots ,
\label{series_Dirac}
\end{eqnarray}
and a similar series for $\psi_R(x)$.
The third term on the right hand side of Equation (\ref{series_Dirac})
can be represented by the Feynman diagram in Figure \ref{Feynman2}.
Note that the propagation of a massive electron can be viewed as
the propagation of its massless Weyl components
$\psi_L$ and $\psi_R$ that forward scatter $0, 1, 2 \cdots$ times
on the vacuum with amplitude $-im$.
Each scattering changes an incoming $\psi_L$ component into
an outgoing $\psi_R$ component, and \textit{vice-versa}.

\begin{figure}
\begin{center}
{\includegraphics{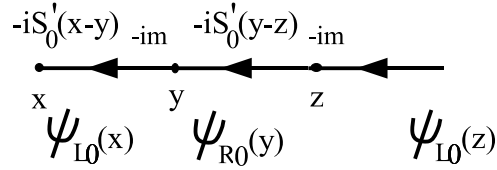}}
\caption{A Feynman diagram of a massless Dirac particle
acquiring mass by forward scattering on the vacuum.}
\label{Feynman2}
\end{center}
\end{figure}

The series (\ref{series_Dirac}) illustrates 
how an electron scatters on vacuum
and acquires mass coupling the $\psi_L$ and $\psi_R$ components
(which is also apparent in (\ref{Dirac})). 
In the limit $m \rightarrow 0$ it is
best to do the integral
\begin{equation}
\int [ -i S'_0(x-y) ] \cdot [ -i S'_0(y-z) ] d^4y = 
S_0(x-z),
\label{integral}
\end{equation}
and recover the series (\ref{series}).

As we shall see, the 
\textquotedblleft{stepping stone}"
picture presented in this Section may be
significantly changed in the standard model
when radiative corrections are taken
into account.

\section{Weyl fields in a box}
Let us use a block-diagonal base such that the first two
components of $\psi$ are $\psi_L$, and the last two components
of $\psi$ are $\psi_R$. In this basis
\begin{equation}
\gamma^0 = \left( 
{\begin{array}{cc}
0 & \sigma^0 \\
\sigma^0 & 0
\end{array} }
\right), 
\gamma^k = \left(  
{\begin{array}{cc}
0 & \sigma^k \\
-\sigma^k & 0
\end{array} }
\right),
\gamma^5 = \left(  
{\begin{array}{cc}
-\sigma^0 & 0 \\
0 & \sigma^0
\end{array} }
\right),
\label{gamma}
\end{equation}
where $\sigma^0$ is the $2 \times 2$ unit matrix, and $\sigma^k$,
with $k = 1, 2, 3$, are the Pauli matrices.

The solution of Equations (\ref{Dirac}) corresponding to
a right handed electron propagating in the $z$-direction is
\begin{equation}
\psi \propto \left(
{\begin{array}{c}
e^{-\phi/2} \\
0 \\
e^{\phi/2} \\
0
\end{array} }
\right)
\times
e^{i(kz - \omega t)},
\label{psi_R}
\end{equation}
where $\omega^2 - k^2 = m^2$, $\omega = m \cosh{\phi}$, 
$k = m \sinh{\phi}$.
The solution corresponding to a left handed
electron propagating in the $-z$-direction is
\begin{equation}
\psi \propto \left(
{\begin{array}{c}
e^{\phi/2} \\
0 \\
e^{-\phi/2} \\
0
\end{array} }
\right)
\times
e^{i(-kz - \omega t)}.
\label{psi_L}
\end{equation}
The limit $m \rightarrow 0$, keeping $\omega \approx k$ finite, 
is singular: the $\psi_L$ and 
$\psi_R$ components become decoupled because $\phi \rightarrow \infty$. 

Consider a box at rest of hight $l$.
A (massless) Weyl-R field propagates up along $z$.
Its $z$-component of spin is $s_z = +\frac{1}{2}$.
A (massless) Weyl-L field is reflected down.
Its $z$-component of spin is $s_z = +\frac{1}{2}$.
Note that angular momentum is conserved in the
reflection. The box has no spin.
We combine (\ref{psi_R}) and (\ref{psi_L}), 
take the limit $\phi \rightarrow \infty$, and change the
notation $\omega \approx k \rightarrow m$. The field 
becomes
\begin{equation}
\propto
\left(
\begin{array}{c}
\exp\{ im(-z-t)\} \\
0 \\
\exp\{ im(z-t)\} \\
0
\end{array}
\right)
\label{mDirac}
\end{equation}
where, in this block diagonal basis, the first two components are
$\psi_L$ propagating down, and the last two components are $\psi_R$
propagating up.

A second observer $(t', x', y', z')$ sees the box moving
in the $z$-direction with velocity $v \equiv \tanh\phi$.
The frequency of the up ward propagating $\psi_R$ field is
Doppler shifted to $\omega' = e^\phi m$, and the time it
takes to traverse the height $l$ of the box is
dilated to $e^\phi l$.
The frequency of the down ward propagating $\psi_L$ field is
Doppler shifted to $\omega' = e^{-\phi} m$, and the time it
takes to traverse the height $l$ of the box is
shortened to $e^{-\phi} l$.
So, the field as viewed by the second observer, is
\begin{equation}
\propto
\left(
\begin{array}{c}
\exp\{-\frac{\phi}{2}\} \exp\{ i e^{-\phi} m (-z' - t')\} \\
0 \\
\exp\{\frac{\phi}{2}\} \exp\{ i e^{\phi} m (z' - t')\} \\
0
\end{array}
\right)
\label{mDirac2}
\end{equation}
as in (\ref{psi_R}) and (\ref{psi_L}) in the zero mass
limit. So we understand why the field transforms as it
does!

Note that the probability of observing the particle moving
up with velocity 1 is $e^\phi / (e^\phi + e^{-\phi})$, and the 
probability of observing the particle moving down 
with velocity -1 is
$e^{-\phi} / (e^\phi + e^{-\phi})$, so the mean
velocity of the particle is $\tanh \phi = v$
as expected.
In the limit $v \rightarrow 1$, $\phi \rightarrow \infty$,
and we are left with only
the field $\psi_R$.
The fields in the box
acquire inertia with mass $m$, as we have explained 
in Section 5.

The model of this section may apply to a quark inside
the proton. If it applies to an electron, then 
a modification of the Dirac equation is needed
(since the Dirac equation only includes forward
scatterings that conserve energy-momentum).
Such a modification may be included in the
standard model as we shall see shortly.

The $\psi_L$ and $\psi_R$ fields of this section 
satisfy the \textit{massless} Dirac equation.
Can the \textit{massive} Dirac equation with mass $m$ be 
applied to the box with Weyl fields as a whole?
If so, the box with massless fields 
$\psi_L$ and $\psi_R$ would be a description of
an electron at a deeper level.
This is similar to neutron diffraction by a 
crystal, described by a wave assigned to the
neutron as a whole, in spite of the internal
wheels and gears of the neutron. As another example,
\textquotedblleft{electrons}" and \textquotedblleft{holes}"
in a semiconductor can be treated as (quasi) 
particles in their own right with their
(effective) masses, or, at a deeper level, 
as electron waves being Bragg-reflected back and forth by
the periodic potential of the crystal.
In fact, the model of standing \textquotedblleft{waves in
a box}" is similar to the standing electron
waves in a crystal when the Bragg condition
is satisfied.

\section{What is the box made of?}
From our preceding discussions on mass and inertia,
it is crucial to conserve energy-momentum as the electron
bounces back and forth across the box. The box
must therefore carry a compensating momentum back and forth.
So we identify the box with a field that carries 
energy-momentum. Since angular momentum is conserved
in the reflection, we choose a scalar field. 
\footnote{A vector field may also be considered. In this
case $\psi_R$ bounces back as $\psi_R$, not as $\psi_L$.
In the standard model the electron mass becomes renormalized
by both the scalar Higgs field and by the gauge boson vector
fields.}
The Feynman diagram of the 
electron bouncing back and forth due to the interaction with
this scalar field $h$ is shown in Figure \ref{box}.
We conclude that the Lagrangian should have terms proportional
to
\begin{equation}
\propto \widetilde{\psi}_L h \psi_R + \widetilde{\psi}_R h \psi_L
\label{vertex}
\end{equation}
for the vertexes, in addition to the terms 
corresponding to the propagators of
the fields $\psi_L$, $\psi_R$ and $h$, if the model of
\textquotedblleft{waves in a box}" is correct.

\begin{figure}
\begin{center}
\scalebox{0.7}
{\includegraphics{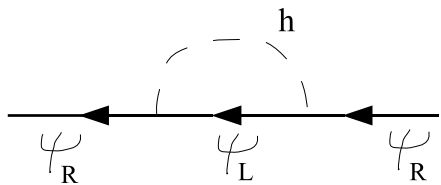}}
\caption{Feynman diagram of an electron in
a \textquotedblleft{box}".}
\label{box}
\end{center}
\end{figure}

\section{Bilinear forms of Weyl fields}
In this Section we use the notation of reference \cite{GMS}
for the irreducible representations of the proper Lorentz
group:
$\left( \frac{1}{2}, -\frac{3}{2} \right)$
is Weyl-R, $\left( \frac{1}{2}, \frac{3}{2} \right)$
is Weyl-L, $\left( 0, 1 \right)$ is scalar,
and $\left( 0, 2 \right)$ is vector.
Every term in the Lagrangian is a scalar with respect
to the proper Lorentz group.
Consider the following
direct products:
\begin{equation}
\left( \frac{1}{2}, \frac{3}{2} \right)
\otimes
\left( \frac{1}{2}, \frac{3}{2} \right)
= \left( 0, 1 \right) \oplus \left( 1, 2 \right),
\label{13}
\end{equation}
\begin{equation}
\left( \frac{1}{2}, -\frac{3}{2} \right)
\otimes
\left( \frac{1}{2}, -\frac{3}{2} \right)
= \left( 0, 1 \right) \oplus \left( 1, -2 \right),
\label{13bar}
\end{equation}
\begin{equation}
\left( \frac{1}{2}, \frac{3}{2} \right)
\otimes
\left( \frac{1}{2}, -\frac{3}{2} \right)
= \left( 0, 2 \right).
\label{4}
\end{equation}
Also, the Hermitian-conjugate of the
matrices of $\left( \frac{1}{2}, -\frac{3}{2} \right)$
carry the representation
$\left( \frac{1}{2}, \frac{3}{2} \right)$, 
and \textit{vice versa}.
So, scalar bilinear forms exist for $\psi_L$
($\psi_R$) that can give chargeless 
Weyl-L (Weyl-R) neutrinos a \textquotedblleft{Majorana mass}".
Similarly, the scalar bilinear form
$\widetilde{\psi}_R \psi_L + \widetilde{\psi}_L \psi_R$
can give the Dirac field a
\textquotedblleft{Dirac mass}".
Finally, the bilinear forms
$\widetilde{\psi}_R \gamma^\mu \psi_R$ and
$\widetilde{\psi}_L \gamma^\mu \psi_L$
are vectors, as needed for terms in the
Lagrangian with first derivatives of the fields,
so, for example, $\widetilde{\psi}_R \gamma^\mu \partial_\mu
\psi_R$ is a scalar.

\begin{figure}
\begin{center}
\scalebox{0.6}
{\includegraphics{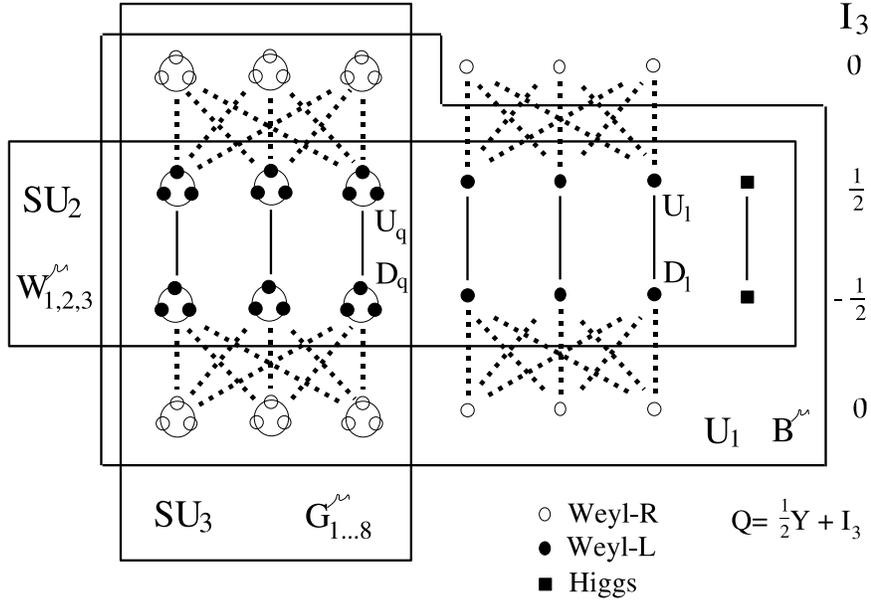}}
\caption{The elementary particles of the standard model.}
\label{SM}
\end{center}
\end{figure}

\section{The standard model}
The standard model is illustrated in Figure \ref{SM}.
It contains Weyl-L fields (black dots), Weyl-R
fields (white dots), and complex Higgs scalars
(squares). The standard model has the following
\textit{global} symmetries:
\begin{itemize}
\item
Invariance with respect to phase
transformations $e^{i Y_j \phi}$ ($U_1$ symmetry)
independently for each field $j$ within the rectangle
\textquotedblleft{$U_1$}". 
\item
Invariance with respect to $SU_2$ 
\textquotedblleft{rotations}" independently among each pair
of fields united by vertical bars inside
the \textquotedblleft{$SU_2$}" rectangle.
All fields within the rectangle
\textquotedblleft{$SU_2$}" carry the representation
$2$ of $SU_2$. Their complex conjugates carry the representation
$\bar{2} = 2$ of $SU_2$.
All fields out of this 
rectangle are singlets of $SU_2$.
\item
Invariance with respect to $SU_3$
transformations 
independently among each triplet
of fields united by circles
inside the \textquotedblleft{$SU_3$}" rectangle.
All fields within the rectangle
\textquotedblleft{$SU_3$}" carry the representation 
\textbf{$3$} of $SU_3$. 
Their complex conjugates carry the representation
$\bar{3}$ of $SU_3$.
All fields out of this
rectangle are singlets of $SU_3$.
\end{itemize}

So far the fields of the model are non-interacting.
To create interactions, \textit{some} of the \textit{global} symmetries
are promoted to \textit{local} symmetries at each space-time point:
\footnote{
Model builders: note that only \textit{some} of the global
symmetries are gauged.
}
\begin{itemize}
\item
\textit{one} $U_1$ transformation with a phase
$\phi$ common to all fields within the rectangle
\textquotedblleft{$U_1$}"
(each of these fields
has its own $Y_j$ quantum number),
\item
\textit{one} $SU_2$ transformation with the three 
generators of $SU_2$ common to all fields within the rectangle
\textquotedblleft{$SU_2$}", and
\item
\textit{one} $SU_3$ transformation with the eight
generators of $SU_3$ common to all fields within the rectangle
\textquotedblleft{$SU_3$}".
\end{itemize}
To achieve the invariance of the model with respect
to these \textit{local} (or \textquotedblleft{gauge}") 
symmetries, it is necessary 
to replace ordinary derivatives $\partial_\mu$
by covariant derivatives 
$D_\mu = \partial_\mu + i g' Y_j B_\mu + 
i g \frac{\vec{\tau}}{2} \cdot \vec{W_\mu}
+ i g_3 T_a G_\mu^a$. 
These covariant derivatives have
one vector gauge field for every generator of the
symmetry group: $B^\mu$ for the generator of $U_1$,
$\vec{W}^\mu$ for the three generators of $SU_2$,
and $G^\mu_1, G^\mu_2, \cdots G^\mu_8$ 
for the 8 generators of $SU_3$.
There are 3 coupling constants:
$g'$ for $U_1$, $g$ for $SU_2$, and
$g_3$ for $SU_3$.
Due to the gauge (and Higgs) fields,
the Lagrangian becomes non-linear, the
fields interact, and \textit{particles} emerge!

The invariance of the Lagrangian with
respect to the transformation $U_1$
implies the conservation of the quantum 
number $Y$ at every vertex of a Feynman
diagram. The invariance with respect
to $SU_2$ implies the conservation of
$I_3$. 
The electric charge is $Q = \frac{1}{2}Y + I_3$.
The invariance with respect to
$SU_3$ implies the conservation of
2 \textquotedblleft{color}"
quantum numbers (because two of the
Gell-Mann $SU_3$ matrices are diagonal).

The symmetries of the standard model prevent
adding mass terms \textquotedblleft{by hand}" 
to the Lagrangian, except to the Higgs 
field.\footnote{Mass terms that break the symmetries
of the theory destroy its renormalizability, 
and no predictions become possible (unless
new physics cuts off the diverging integrals).}
All other particles acquire mass if the ground
state of the Higgs field 
acquires a vacuum expectation value and
\textquotedblleft{hides}" the $SU_2$ symmetry.
Then the interactions with the vacuum expectation value
of the Higgs field pairs up Weyl-L and Weyl-R fields
giving mass to the resulting Dirac fields.
The Weyl fields that are paired up
are \textquotedblleft{rotated}" among the three families,
as indicated in Figure \ref{SM} by the matrices
$U_q$, $D_q$, $U_l$, $D_l$, and by the dotted lines.
The unitary Cabibbo-Kobayashi-Maskawa matrix is
$U^\dag_q D_q$. There is a similar matrix 
$U^\dag_l D_l$ mixing
families in the lepton sector. No other
rotations among families have experimental
consequences. The mass of the Dirac fields
can not be calculated in the standard model
because their \textquotedblleft{Yukawa coupling}"
to the Higgs are put in by hand, and also because the
masses become renormalized by radiative corrections.
So, in the end, \textit{the experimental masses 
of quarks and leptons must be put
in by hand at each order in perturbation theory:}
at least for the time being, 
\textit{they are not calculable even in principle!}

The real magic of the Higgs mechanism is how it gives
mass to the $W^\pm$ and $Z$ bosons, leaving the photon massless.
The vacuum expectation value $v$ of the Higgs
field gives the $W^\pm$ a \textquotedblleft{tree level}"
mass $vg/2$,
and the $Z$ a \textquotedblleft{tree level}" 
mass $v(g^2 + g'^2)^{1/2}/2$.
The complex Higgs $SU_2$ doublet has four amplitudes.
Three of them can be \textquotedblleft{gauged away}"
and become the longitudinal polarizations needed by
$W^\pm$ and $Z$ to become massive. 
Diagrams that contain these longitudinal
polarizations diverge. These divergences are canceled,
order by order in the perturbation expansion, by the only 
remaining amplitude $h$ of the Higgs field. It is
magic!

\section{Mass revisited}
The term in the standard model Lagrangian that gives mass
to the electron is \cite{Itzykson_Zuber}
\begin{equation}
-G_e \frac{v + h}{\sqrt{2}} 
\left( \widetilde{\psi}_R \psi_L + 
\widetilde{\psi}_L \psi_R \right),
\label{mass_electron}
\end{equation}
where $G_e$ is a
dimensionless \textquotedblleft{Yukawa coupling}"
put in by hand.
So the vacuum expectation value $v$ of the Higgs
field gives the electron a \textquotedblleft{tree level}"
mass $m_e = G_e v/\sqrt{2}$. 
Note in (\ref{mass_electron}) that
the Lagrangian indeed contains a
term (\ref{vertex}). So the standard model contains
loops as shown in Figure \ref{box}. Summing a series
with $0, 1, 2 \cdots$ loops we obtain a propagator 
with the renormalized mass
\begin{equation}
m_e = \frac{G_e}{\sqrt{2}}
\left(v + \frac{G_e}{\sqrt{2}} A
+ \cdots \right),
\label{renormalized_mass}
\end{equation}
where the factor $A$ has the form
\begin{equation}
A = -i \int{ 
\frac{d^4 k}{(2\pi)^4}
\frac{i}{k_\mu k^\mu - m_h^2}
\frac{i}{\gamma^\mu (p_\mu + k_\mu) - \frac{G_e v}{\sqrt{2}}}
}.
\label{A}
\end{equation}
The contributions of 
higher order primitive graphs are indicated by dots in 
(\ref{renormalized_mass}).

In the standard model,
the integral (\ref{A}) diverges (apparently) linearly 
as the regularization is turned off. In this
case, the second term in the parenthesis of 
(\ref{renormalized_mass}) diverges, keeping
$m_e = \frac{1}{2}G_e^2 A$ finite.
However, the standard model is not expected to
be the final theory of nature. New physics will
emerge at the Planck scale, or grand unified scale,
or even lower, and effectively cuts off the integral
(\ref{A}). 

We then have two alternatives.
In the first alternative, the term
with $v$ dominates in (\ref{renormalized_mass}), and the
electron acquires mass by forward scattering on the vacuum
as explained in Section 8. 
This is the \textquotedblleft{stepping stone}" model.
In the second alternative, the terms 
$\frac{G_e}{\sqrt{2}} A + \cdots$ dominate, 
and we obtain the model of Weyl-L and
Weyl-R fields reflecting back and forth in a box made of
the Higgs field as described in Sections 9 and 10.
This is the \textquotedblleft{waves in a box}" model.
The two alternatives
are shown schematically in Figure \ref{mass_model}.
Which alternative has nature chosen? Or has
nature chosen a combination of both? 

\begin{figure}
\begin{center}
\scalebox{0.5}
{\includegraphics{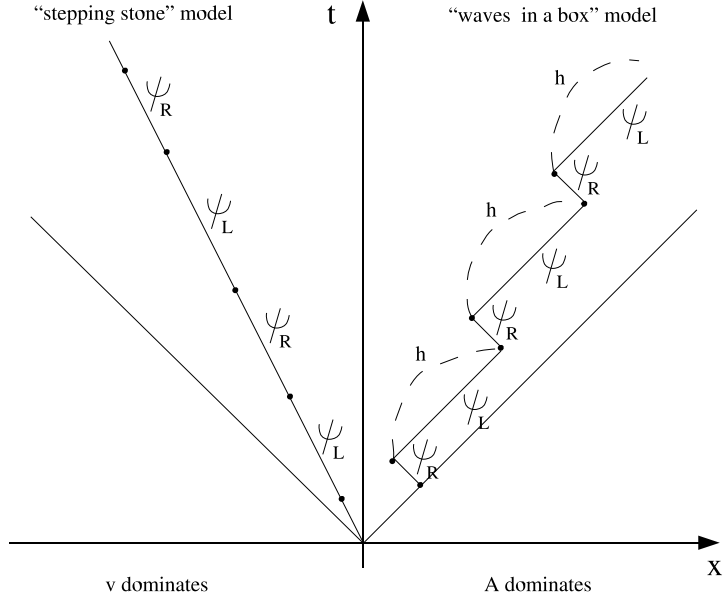}}
\caption{Two alternatives to acquire mass.}
\label{mass_model}
\end{center}
\end{figure}

\section{Experimental consequences}

The proton can be viewed as ultrarelativistic
quarks in a box made of gluons. The radius of
the box has been measured, for example, with
$\gamma p \rightarrow \gamma p$ scattering.
Note that the proton acquires mass independently
of the vacuum expectation value of the
Higgs field, and this mass can be calculated
with lattice quantum chromodynamics.

Note that scattering with momentum transfer
$\sqrt{ | t | } \lesssim 4.6/R$
probes the radius $R$ of the box as a whole,
while scattering with $\sqrt{ | t | } \gtrsim 4.6/R$
probes its constituents.

The case of the electron is more
uncertain. Apparently, the electron
acquires mass by the 
\textquotedblleft{stepping stone}"
mechanism. If the electron can be
considered as
\textquotedblleft{waves in a box}",
then the box has size
$\approx \hbar/(m c)$, the field
$h$ would have to be nearly massless,
and the size of the box would cause
the Darwin shift of the energy levels
of hydrogen.

Many questions remain.
The reason for an effective, low energy,
theory to be renormalizable is to
decouple it from the high energy
theory. But, in the standard model, the
Higgs sector does not decouple.
What are the consequences on $e^+ e^-$
scattering of Feynman graphs with
Higgs in a loop? If new physics
cuts off otherwise divergent integrals
such as (\ref{A}), what are the experimental
consequences? Can we constrain new physics
this way?

\section{Conclusions}
I have tried to understand the origin of mass
and have failed miserably (which may be a step in the
right direction). 
Nature, as we currently understand it,
is described by a perturbation expansion of Feynman diagrams.
We subscribe to the belief that \textquotedblleft{the
diagrams contain more truth than the underlying formalism}"
['t Hooft and Veltman (1973)]. 
We can not calculate the electron mass: we must introduce
a \textquotedblleft{Yukawa coupling}" by hand 
\textit{at each order
in the perturbation expansion}.
So, at our present level of understanding,
\textit{it appears hopeless to calculate the masses of 
quarks and leptons}:
we must take their values from experiment.
On the other hand, we can calculate the mass of baryons and
mesons using lattice quantum chromodynamics.

Early in this paper we arrived at a beautiful insight.
We considered \textit{classical} waves that interact. We found that
the condition that the waves add up in phase 
at the receiver has the form
\begin{eqnarray}
n \omega_1 + m \omega_2 + \cdots & = & 0, \nonumber \\
n \vec{k}_1 + m \vec{k}_2 + \cdots & = & 0,
\label{particles}
\end{eqnarray}
where the powers $n, m \cdots$ of the Taylor expansion
of the non-linear interaction,
are integers. Note that the 
dispersion occurs \textit{as if the waves were composed of
discrete packets} of frequency-wavevector $( \omega_j, \vec{k}_j )$.
The packets are called \textquotedblleft{particles}",
and $( E_j, \vec{p}_j ) \equiv ( \hbar \omega_j, \hbar \vec{k}_j )$
is the \textquotedblleft{energy-momentum}" of particle $j$. According to
(\ref{particles}), energy-momentum is conserved (with a precision
determined by the uncertainty principle, i.e. by 
the size of the region of space-time over which the
interference builds up). Note
that the concept of particle,
and the Planck and De Broglie
relations, have emerged from the interaction
of \textit{classical} waves!

In semiconductor physics we can consider quasi-particles called
\textquotedblleft{electrons}" and \textquotedblleft{holes}"
with their effective masses, or, at a deeper level, we can 
consider electrons being Bragg-reflected back and forth by the
periodic potential of the crystal. Similarly, we can
consider the massive Dirac electron, or, at a deeper level,
we can perhaps consider massless Weyl-L and Weyl-R fields being reflected
back and forth by emitting and absorbing virtual scalar
Higgs particles. Similarly, we can perhaps consider massive
gauge bosons as massless bosons being reflected back and
forth by emitting and absorbing virtual scalar
Higgs particles. These reflections give the gauge bosons
their longitudinal polarization (as reflections in a 
waveguide give the photon a mass and a longitudinal
polarization). 
Which model best describes nature,
\textquotedblleft{waves in a box}" or
\textquotedblleft{stepping stone}", 
or a combination of both,
will depend on which term in (\ref{renormalized_mass}) 
dominates.

The perturbation expansion of Feynman diagrams has
all imaginable diseases. We may hope that an expansion
is possible around a less singular starting point.
For some applications it may prove convenient to
replace massive fermions and bosons by the corresponding
\textquotedblleft{waves in a box}".

In the process of trying to understand mass, 
I have arrived at the following view of the
standard model. Quarks and leptons are not 
elementary at all. An electron is a messy composite
of a superposition of three Weyl-L fields, 
a Weyl-R field and a Higgs field coupled 
together by a
Yukawa coupling, and all of them are 
coupled to gauge fields by gauge couplings,
which also interact with themselves and
everything else. The energy of this mess
has inertia, as all energy does, and it is that mess 
that determines the mass of the electron.
So the next level in our understanding of Nature,
from atoms to nuclei and electrons,
to baryons and mesons, to quarks and leptons, are
the dots shown in Figure \ref{SM}, and these have
no mass (except the Higgs?).

\end{document}